\documentstyle[preprint,aps]{revtex}
\newcommand{\be}{\begin{equation}}
\newcommand{\ee}{\end{equation}}
\newcommand{\ba}{\begin{eqnarray}}
\newcommand{\ea}{\end{eqnarray}}
\begin{document}
\preprint{HYUPT 0819/96}
\draft
\title{Asymptotic Conformal Invariance of SU(2) and Standard Models
in Curved Space-time}
\author{Youngsoo Yoon and Yongsung Yoon}
\address{Department of Physics, Hanyang University, 
Seoul, 133-791, Korea}
\maketitle
\begin{abstract}
The asymptotic conformal invariance of some SU(2) model and Standard Model 
in curved space-time are investigated. We have examined the conditions for 
asymptotic conformal invariance for these models numerically.
\end{abstract}
\pacs{04.62.+v, 04.60.-m}

\newpage
\pagenumbering{arabic}\setcounter{page}{1}

\noindent
\section{Introduction}

The asymptotic behavior of quantum field theory(QFT) in curved space-time is
quite important in the early Universe considerations. The typical action for
the arbitrary renormalizable theory(including scalar sector) in curved
space-time has the following form \cite{kn:1};
\ba
S = \int d^nx \sqrt{-g} ( L_{matter} + \frac{1}{2} \xi R \varphi^2 +L_{ext} )
\label{eq:1}
\ea
where $L_{matter}$ is the same Lagrangian as typical for flat space-time
(with changes $\partial_\mu \rightarrow \nabla_\mu , \gamma^\mu \rightarrow
\gamma^\mu (x)$), and $L_{ext}$ is the Lagrangian  for external fields
\cite{kn:2}.
It is quite well-known that the asymptotic behavior of an arbitrary QFT is
best investigated through renormalization group(RG) where coupling constants
are getting effective coupling constants.

In the action Eq.(\ref{eq:1}) the new scalar-gravitational coupling
constant $\xi$ appears
(if compared with flat space-time). The asymptotic behavior of this
effective coupling in non-abelian asymptotically free gauge theories has been
first investigated in Ref.\cite{kn:2}. In these works, Buchbinder and Odintsov
has shown that at high energies (strong curvature) the beautiful phenomenon,
asymptotic conformal invariance, may be realized. It means that
$ \xi(t) \rightarrow \frac{1}{6} $ as $ t \rightarrow \infty $ where $t$
is RG parameter, {\it i.e.} in matter sector theory is trying 
to become conformally
invariant at high energies. (Later on, the other types of $\xi(t)$ behavior
have been found \cite{kn:2,kn:3}
like $\mid \xi(t) \mid  \rightarrow \infty$ or $\xi(t) = \xi$ for a theory
\cite{kn:1,kn:2,kn:3,kn:4})

The value of $\xi$ is important in realization of different types of
inflationary Universe. For example, for some inflationary models $\xi$ should
be very large. 

In the present letter, we study the behavior of $\xi(t)$ in SU(2) gauge
theories with spinors and scalars introduced in Ref.\cite{kn:5} in flat
space-time. However, we study its behavior not in asymptotically free regime
on special solutions of
RG as it has been done in Ref.\cite{kn:1,kn:2} but in general solutions
(hence, numerically).
We find that the theory is still asymptotically conformally invariant. In the
section 3, we discuss the RG behavior of $\xi(t)$ numerically in the
Standard Model(SM). It is found that SM does not have asymptotic conformal
invariance in general except for a special case, $\xi_0 = \frac{1}{6}$.

\section{SU(2) Model in Curved Space-Time}

In curved space-time, the SU(2) gauge theory with scalars and spinors
\cite{kn:1} had been investigated when the gauge coupling $g$, Yukawa
coupling $h$ and quartic scalar coupling $f$ are asymptotically free
on special solutions of RG equations \cite{kn:5}.
We have investigated the behavior of $\xi$ in general.\\
The form of the SU(2) gauge theory in curved space-time is \cite{kn:1} \\
\ba
S &=& \int d^nx \sqrt{-g} [ L_{YM} + \sum_{k=1}^{m} i \psi_{(k)}^a
\gamma^\mu(x) D_\mu^{ab} \psi_{(k)}^b \nonumber \\
&+& \frac{1}{2} (D_\mu^{ab} \varphi_b )^2
- i h \epsilon^{acb} \psi_{(k)}^a \psi_{(k)}^b \varphi_c -
\frac{f}{4!} (\varphi_a^2 )^2
+ \frac{1}{2} \xi R \varphi^2 ] , \\
L_{YM} &=& - \frac{1}{4} ( \nabla_\mu A_\nu^a - \nabla_\nu
A_\mu^a
+ g \epsilon^{abc} A_\mu^b A_\nu^c )^2 \nonumber.
\ea \\
This action contains scalars $\varphi^a$, spinors $\psi_k^a$ belonging to
adjoint representation of the gauge group and gauge fields
$A_\mu^a;~ a=(1,2,3),~k=(1,2,3,...,m),~ m$ is the number of spinor multiplets.
$D_\mu^{ab}$ is the general covariant derivative, and it is defined as
follows;
\ba
D_\mu^{ab} \varphi_b &=& \nabla_\mu \delta^{ab} \varphi_b
+ i g \epsilon^{acb} A_\mu^c \varphi_b ,\nonumber \\
D_\mu^{ab} \psi_b &=& \nabla_\mu \delta^{ab} \psi_b
+ i g \epsilon^{acb} A_\mu^c \psi_b
+ \frac{1}{2} \omega_\mu^{\alpha\beta} \sigma_{\alpha\beta} \psi_b
\ea \\
where the matrices $\sigma_{\alpha\beta}$ are given by the relation
\ba
\sigma_{\alpha\beta} = \frac{1}{2} (\gamma_\alpha
\gamma_\beta - \gamma_\beta \gamma_\alpha), \nonumber
\ea \\
and $\omega_\mu^{\alpha\beta}$
is the spinor connection which satisfies the torsionless condition;
\ba
\partial_\mu e_\nu^\alpha -\partial_\nu e_\mu^\alpha + e_{\mu\beta}
\omega_\nu^{\alpha\beta} - e_{\nu\beta} \omega_\mu^{\alpha\beta} = 0.
\ea \\
From the above, following four equations are obtained using the Schwinger-De
Witt technique \cite{kn:1,kn:6};
\ba
(4 \pi)^2 \frac{dg^2}{dt} &=& - ( 14 - \frac{16}{3} m ) g^4 , \\
(4 \pi)^2 \frac{dh^2}{dt} &=&  16 h^4 - 24 g^2 h^2 , \label{eq:6} \\
(4 \pi)^2 \frac{df}{dt} &=& \frac{11}{3} f^2 - 24 g^2 f + 72 g^4
+ 16fh^2 - 96 h^4 , \label{eq:7} \\
(4 \pi)^2 \frac{d\xi}{dt} &=& (\xi - \frac{1}{6})( \frac{5}{3} f + 8 h^2 -
12g^2 ) . \label{eq:8}
\ea
The special solutions of $g^2(t)$, $h^2(t)$ and $f^2(t)$ are \cite{kn:5}
\ba
g^2 (t) &=& \frac{g_0^2}{1 + c g_0^2 t}, ~~~
c \equiv \frac{1}{16 \pi^2} ( 14 - \frac{16}{3} m ) , \\
h^2 (t) &=& ( \frac{5}{8} + \frac{1}{3} m ) g^2 (t) , \label{eq:10} \\
f^2 (t) &=& ( \frac{120}{11} m + \frac{32}{11} m^2
- \frac{207}{22} ) g^4 (t) .
\ea
These solutions are asymptotically free for $m=1,2$.\\
With the above asymptotically free solutions, the solution of $\xi$ is
\be
\xi = (\xi_0 - \frac{1}{6}) ( 1 + cg_0^2 t )^\frac{B}{c} + \frac{1}{6} ,
\ee
where
\be
B \equiv \frac{1}{16\pi^2} [\frac{5}{3}
\sqrt{\frac{120}{11} m + \frac{32}{11} m^2 -\frac{207}{22}}
+ 8 (\frac{5}{8} + \frac{1}{3} m ) -12 ].
\ee
For $m=1$, $\xi \rightarrow \frac{1}{6}$ as $ t \rightarrow \infty $  
($\frac{B}{c} \approx -0.38 < 0$). \\
For $m=2$, $\xi \rightarrow \infty$ as $t \rightarrow \infty $
($\frac{B}{c} \approx 7.58 > 0$). \\
This case has been investigated in Ref.\cite{kn:2} already.

However we investigated the RG-solution of $\xi$ for
the general solutions of the other couplings ($h$ and $f$).
The general solution of the $h^2(t)$ is
\ba
h^2(t) &=& g^2(t)  \frac{(g_0^2 )^{-\frac{a}{c}} h_0^2
 a \pi^2 }{ ( g_0^2 )^{-\frac{a}{c}} h_0^2 + ( g^2(t) )^{-\frac{a}{c}}
[a \pi^2  g_0^2 - h_0^2] } , \\
a &\equiv& \frac{1}{\pi^2} ( \frac{5}{8} + \frac{1}{3} m ) . \nonumber
\ea
It is the unique solution(Eq.(\ref{eq:6}) can be linearized). The $h^2 (t)$ has
asymptotically free solution only if $m=1,2$.
For $m \geq 3$, the $h^2 (t)$ becomes negative.
The initial value of $h_0$ is crucial.
Due to the positivity condition,
only for $ h_0^2 \leq ( \frac{5}{8} + \frac{1}{3} m ) g_0^2 \equiv h_s^2$,
the solution is meaningful.
In the case of equality, the general solution is reduced to the special
solution of Eq.(\ref{eq:10}) [Fig.1].

The solution of $f$ has two types.
If the solution of $h^2 (t)$ is the special one({\it i.e.} $h_0^2 =h_s^2 $),
we have the analytical solution;
\be
 f(t) = g^2 (t) A \frac{Q_0 + A + \lambda ( Q_0 - A ) }{Q_0 + A
- \lambda ( Q_0 - A ) } ,
\ee
where
$~~~~~Q_0 \equiv \frac{f_0}{g_0^2},
~~~A \equiv \sqrt{\frac{24}{11} ( 5 m + \frac{4}{3} m^2 - \frac{69}{16})},
~~~\lambda \equiv ( 1 + c g_0^2 t )^{\frac{11A}{24 \pi^2 c}} .$ \\ ~\\

\setlength{\unitlength}{0.240900pt}
\ifx\plotpoint\undefined\newsavebox{\plotpoint}\fi
\sbox{\plotpoint}{\rule[-0.200pt]{0.400pt}{0.400pt}}%
\special{em:linewidth 0.4pt}%
\\ ~\\

The solution becomes the special solution Eq.(\ref{eq:10}) for $Q_0 = A$,
{\it i.e.} $f_0 = A g_0^2 \equiv f_s $.\\
For $Q_0 > A$, ({\it i.e.} $f_0 > f_s$), the solution has a singularity.
For $m=1$ and $2$, $\lambda \rightarrow \infty$ as $t \rightarrow \infty$
because $c$ and $\frac{A}{c}$ are positive. 
In this case, we have an approximate solution of $f(t)$ as follow;
\be
f (t) \approx - g^2 (t) \sqrt{\frac{24}{11} ( 5 m +
\frac{4}{3} m^2 - \frac{69}{16})} ,
\ee
which is negative even though very small at large $t$ [Fig.2].
In the case of $h_0^2 = h_s^2$, $f(t)$ changes its behavior near $f_0 = f_s$
sensitively.
For $ h_0^2 < h_s^2 $, it is very hard to deal with the Eq.(\ref{eq:7})
analytically.
From the definition of $a$ and $c$, $\frac{a}{c} = - \frac{23}{13}$
for $m=1$ and $\frac{a}{c} = - \frac{31}{5}$ for $m=2$. \\ ~\\

\setlength{\unitlength}{0.240900pt}
\ifx\plotpoint\undefined\newsavebox{\plotpoint}\fi
\sbox{\plotpoint}{\rule[-0.200pt]{0.400pt}{0.400pt}}%
\special{em:linewidth 0.4pt}%
\begin{picture}(1500,900)(0,0)
\font\gnuplot=cmr10 at 10pt
\gnuplot
\put(176,113){\special{em:moveto}}
\put(1436,113){\special{em:lineto}}
\put(176,113){\special{em:moveto}}
\put(176,877){\special{em:lineto}}
\put(176,113){\special{em:moveto}}
\put(196,113){\special{em:lineto}}
\put(1436,113){\special{em:moveto}}
\put(1416,113){\special{em:lineto}}
\put(154,113){\makebox(0,0)[r]{0}}
\put(176,209){\special{em:moveto}}
\put(196,209){\special{em:lineto}}
\put(1436,209){\special{em:moveto}}
\put(1416,209){\special{em:lineto}}
\put(154,209){\makebox(0,0)[r]{0.5}}
\put(176,304){\special{em:moveto}}
\put(196,304){\special{em:lineto}}
\put(1436,304){\special{em:moveto}}
\put(1416,304){\special{em:lineto}}
\put(154,304){\makebox(0,0)[r]{1}}
\put(176,400){\special{em:moveto}}
\put(196,400){\special{em:lineto}}
\put(1436,400){\special{em:moveto}}
\put(1416,400){\special{em:lineto}}
\put(154,400){\makebox(0,0)[r]{1.5}}
\put(176,495){\special{em:moveto}}
\put(196,495){\special{em:lineto}}
\put(1436,495){\special{em:moveto}}
\put(1416,495){\special{em:lineto}}
\put(154,495){\makebox(0,0)[r]{2}}
\put(176,591){\special{em:moveto}}
\put(196,591){\special{em:lineto}}
\put(1436,591){\special{em:moveto}}
\put(1416,591){\special{em:lineto}}
\put(154,591){\makebox(0,0)[r]{2.5}}
\put(176,686){\special{em:moveto}}
\put(196,686){\special{em:lineto}}
\put(1436,686){\special{em:moveto}}
\put(1416,686){\special{em:lineto}}
\put(154,686){\makebox(0,0)[r]{3}}
\put(176,782){\special{em:moveto}}
\put(196,782){\special{em:lineto}}
\put(1436,782){\special{em:moveto}}
\put(1416,782){\special{em:lineto}}
\put(154,782){\makebox(0,0)[r]{3.5}}
\put(176,877){\special{em:moveto}}
\put(196,877){\special{em:lineto}}
\put(1436,877){\special{em:moveto}}
\put(1416,877){\special{em:lineto}}
\put(154,877){\makebox(0,0)[r]{$f(t)$}}
\put(176,113){\special{em:moveto}}
\put(176,133){\special{em:lineto}}
\put(176,877){\special{em:moveto}}
\put(176,857){\special{em:lineto}}
\put(176,68){\makebox(0,0){0}}
\put(450,113){\special{em:moveto}}
\put(450,133){\special{em:lineto}}
\put(450,877){\special{em:moveto}}
\put(450,857){\special{em:lineto}}
\put(450,68){\makebox(0,0){50}}
\put(724,113){\special{em:moveto}}
\put(724,133){\special{em:lineto}}
\put(724,877){\special{em:moveto}}
\put(724,857){\special{em:lineto}}
\put(724,68){\makebox(0,0){100}}
\put(998,113){\special{em:moveto}}
\put(998,133){\special{em:lineto}}
\put(998,877){\special{em:moveto}}
\put(998,857){\special{em:lineto}}
\put(998,68){\makebox(0,0){150}}
\put(1272,113){\special{em:moveto}}
\put(1272,133){\special{em:lineto}}
\put(1272,877){\special{em:moveto}}
\put(1272,857){\special{em:lineto}}
\put(1272,68){\makebox(0,0){200}}
\put(176,113){\special{em:moveto}}
\put(1436,113){\special{em:lineto}}
\put(1436,877){\special{em:lineto}}
\put(176,877){\special{em:lineto}}
\put(176,113){\special{em:lineto}}
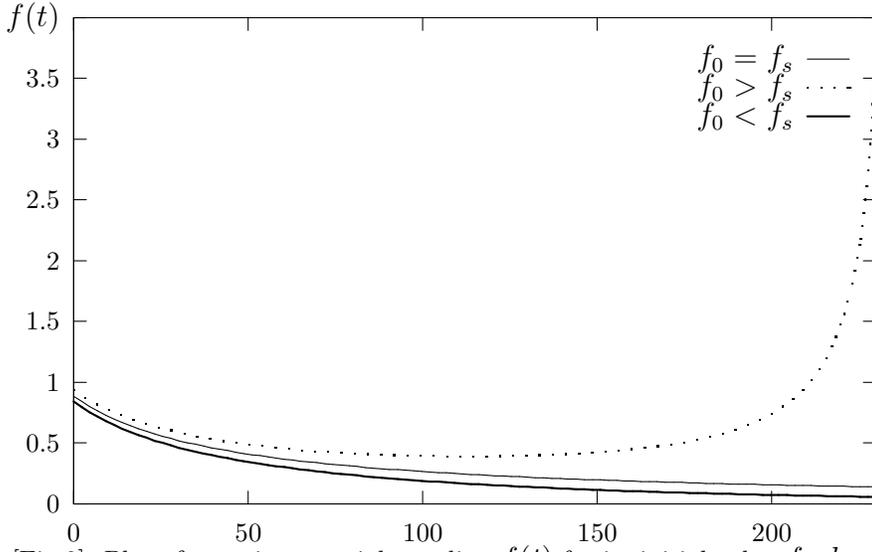
\put(806,23){\makebox(0,0){[Fig.2]. Plot of quartic potential coupling
$f(t)$ for its initial value $f_0$, $h_0 = h_s$.}}
\put(1306,812){\makebox(0,0)[r]{$f_0 = f_s$}}
\put(1328,812){\special{em:moveto}}
\put(1394,812){\special{em:lineto}}
\put(176,282){\special{em:moveto}}
\put(189,274){\special{em:lineto}}
\put(201,266){\special{em:lineto}}
\put(214,259){\special{em:lineto}}
\put(227,252){\special{em:lineto}}
\put(240,246){\special{em:lineto}}
\put(252,241){\special{em:lineto}}
\put(265,236){\special{em:lineto}}
\put(278,231){\special{em:lineto}}
\put(291,227){\special{em:lineto}}
\put(303,223){\special{em:lineto}}
\put(316,219){\special{em:lineto}}
\put(329,216){\special{em:lineto}}
\put(341,212){\special{em:lineto}}
\put(354,209){\special{em:lineto}}
\put(367,207){\special{em:lineto}}
\put(380,204){\special{em:lineto}}
\put(392,201){\special{em:lineto}}
\put(405,199){\special{em:lineto}}
\put(418,197){\special{em:lineto}}
\put(431,194){\special{em:lineto}}
\put(443,192){\special{em:lineto}}
\put(456,190){\special{em:lineto}}
\put(469,189){\special{em:lineto}}
\put(481,187){\special{em:lineto}}
\put(494,185){\special{em:lineto}}
\put(507,183){\special{em:lineto}}
\put(520,182){\special{em:lineto}}
\put(532,180){\special{em:lineto}}
\put(545,179){\special{em:lineto}}
\put(558,178){\special{em:lineto}}
\put(571,176){\special{em:lineto}}
\put(583,175){\special{em:lineto}}
\put(596,174){\special{em:lineto}}
\put(609,173){\special{em:lineto}}
\put(621,172){\special{em:lineto}}
\put(634,170){\special{em:lineto}}
\put(647,169){\special{em:lineto}}
\put(660,168){\special{em:lineto}}
\put(672,167){\special{em:lineto}}
\put(685,167){\special{em:lineto}}
\put(698,166){\special{em:lineto}}
\put(711,165){\special{em:lineto}}
\put(723,164){\special{em:lineto}}
\put(736,163){\special{em:lineto}}
\put(749,162){\special{em:lineto}}
\put(761,162){\special{em:lineto}}
\put(774,161){\special{em:lineto}}
\put(787,160){\special{em:lineto}}
\put(800,159){\special{em:lineto}}
\put(812,159){\special{em:lineto}}
\put(825,158){\special{em:lineto}}
\put(838,157){\special{em:lineto}}
\put(851,157){\special{em:lineto}}
\put(863,156){\special{em:lineto}}
\put(876,156){\special{em:lineto}}
\put(889,155){\special{em:lineto}}
\put(901,154){\special{em:lineto}}
\put(914,154){\special{em:lineto}}
\put(927,153){\special{em:lineto}}
\put(940,153){\special{em:lineto}}
\put(952,152){\special{em:lineto}}
\put(965,152){\special{em:lineto}}
\put(978,151){\special{em:lineto}}
\put(991,151){\special{em:lineto}}
\put(1003,151){\special{em:lineto}}
\put(1016,150){\special{em:lineto}}
\put(1029,150){\special{em:lineto}}
\put(1041,149){\special{em:lineto}}
\put(1054,149){\special{em:lineto}}
\put(1067,148){\special{em:lineto}}
\put(1080,148){\special{em:lineto}}
\put(1092,148){\special{em:lineto}}
\put(1105,147){\special{em:lineto}}
\put(1118,147){\special{em:lineto}}
\put(1131,147){\special{em:lineto}}
\put(1143,146){\special{em:lineto}}
\put(1156,146){\special{em:lineto}}
\put(1169,145){\special{em:lineto}}
\put(1181,145){\special{em:lineto}}
\put(1194,145){\special{em:lineto}}
\put(1207,144){\special{em:lineto}}
\put(1220,144){\special{em:lineto}}
\put(1232,144){\special{em:lineto}}
\put(1245,144){\special{em:lineto}}
\put(1258,143){\special{em:lineto}}
\put(1271,143){\special{em:lineto}}
\put(1283,143){\special{em:lineto}}
\put(1296,142){\special{em:lineto}}
\put(1309,142){\special{em:lineto}}
\put(1321,142){\special{em:lineto}}
\put(1334,142){\special{em:lineto}}
\put(1347,141){\special{em:lineto}}
\put(1360,141){\special{em:lineto}}
\put(1372,141){\special{em:lineto}}
\put(1385,141){\special{em:lineto}}
\put(1398,140){\special{em:lineto}}
\put(1411,140){\special{em:lineto}}
\put(1423,140){\special{em:lineto}}
\put(1436,140){\special{em:lineto}}
\put(1306,767){\makebox(0,0)[r]{$f_0 > f_s$}}
\multiput(1328,767)(20.756,0.000){4}{\usebox{\plotpoint}}
\put(1394,767){\usebox{\plotpoint}}
\put(176,291){\usebox{\plotpoint}}
\put(176.00,291.00){\usebox{\plotpoint}}
\put(193.57,279.95){\usebox{\plotpoint}}
\put(211.41,269.39){\usebox{\plotpoint}}
\multiput(214,268)(18.845,-8.698){0}{\usebox{\plotpoint}}
\put(230.18,260.53){\usebox{\plotpoint}}
\put(249.17,252.18){\usebox{\plotpoint}}
\multiput(252,251)(19.372,-7.451){0}{\usebox{\plotpoint}}
\put(268.60,244.89){\usebox{\plotpoint}}
\put(288.43,238.79){\usebox{\plotpoint}}
\multiput(291,238)(19.690,-6.563){0}{\usebox{\plotpoint}}
\put(308.28,232.78){\usebox{\plotpoint}}
\put(328.51,228.11){\usebox{\plotpoint}}
\multiput(329,228)(20.136,-5.034){0}{\usebox{\plotpoint}}
\put(348.68,223.23){\usebox{\plotpoint}}
\multiput(354,222)(20.224,-4.667){0}{\usebox{\plotpoint}}
\put(368.93,218.70){\usebox{\plotpoint}}
\put(389.42,215.43){\usebox{\plotpoint}}
\multiput(392,215)(20.224,-4.667){0}{\usebox{\plotpoint}}
\put(409.75,211.27){\usebox{\plotpoint}}
\put(430.37,209.05){\usebox{\plotpoint}}
\multiput(431,209)(20.473,-3.412){0}{\usebox{\plotpoint}}
\put(450.86,205.79){\usebox{\plotpoint}}
\multiput(456,205)(20.694,-1.592){0}{\usebox{\plotpoint}}
\put(471.49,203.59){\usebox{\plotpoint}}
\put(492.08,201.15){\usebox{\plotpoint}}
\multiput(494,201)(20.694,-1.592){0}{\usebox{\plotpoint}}
\put(512.77,199.56){\usebox{\plotpoint}}
\multiput(520,199)(20.473,-3.412){0}{\usebox{\plotpoint}}
\put(533.34,196.90){\usebox{\plotpoint}}
\put(554.03,195.31){\usebox{\plotpoint}}
\multiput(558,195)(20.756,0.000){0}{\usebox{\plotpoint}}
\put(574.76,194.69){\usebox{\plotpoint}}
\put(595.45,193.04){\usebox{\plotpoint}}
\multiput(596,193)(20.694,-1.592){0}{\usebox{\plotpoint}}
\put(616.17,192.00){\usebox{\plotpoint}}
\multiput(621,192)(20.694,-1.592){0}{\usebox{\plotpoint}}
\put(636.88,190.78){\usebox{\plotpoint}}
\put(657.60,190.00){\usebox{\plotpoint}}
\multiput(660,190)(20.684,-1.724){0}{\usebox{\plotpoint}}
\put(678.32,189.00){\usebox{\plotpoint}}
\multiput(685,189)(20.756,0.000){0}{\usebox{\plotpoint}}
\put(699.07,188.92){\usebox{\plotpoint}}
\put(719.79,188.00){\usebox{\plotpoint}}
\multiput(723,188)(20.756,0.000){0}{\usebox{\plotpoint}}
\put(740.54,188.00){\usebox{\plotpoint}}
\multiput(749,188)(20.684,-1.724){0}{\usebox{\plotpoint}}
\put(761.26,187.00){\usebox{\plotpoint}}
\put(782.01,187.00){\usebox{\plotpoint}}
\multiput(787,187)(20.756,0.000){0}{\usebox{\plotpoint}}
\put(802.77,187.00){\usebox{\plotpoint}}
\put(823.52,187.00){\usebox{\plotpoint}}
\multiput(825,187)(20.756,0.000){0}{\usebox{\plotpoint}}
\put(844.26,187.48){\usebox{\plotpoint}}
\multiput(851,188)(20.756,0.000){0}{\usebox{\plotpoint}}
\put(865.00,188.00){\usebox{\plotpoint}}
\put(885.75,188.00){\usebox{\plotpoint}}
\multiput(889,188)(20.684,1.724){0}{\usebox{\plotpoint}}
\put(906.47,189.00){\usebox{\plotpoint}}
\multiput(914,189)(20.694,1.592){0}{\usebox{\plotpoint}}
\put(927.18,190.00){\usebox{\plotpoint}}
\put(947.91,190.66){\usebox{\plotpoint}}
\multiput(952,191)(20.694,1.592){0}{\usebox{\plotpoint}}
\put(968.61,192.00){\usebox{\plotpoint}}
\put(989.34,192.87){\usebox{\plotpoint}}
\multiput(991,193)(20.684,1.724){0}{\usebox{\plotpoint}}
\put(1010.02,194.54){\usebox{\plotpoint}}
\multiput(1016,195)(20.694,1.592){0}{\usebox{\plotpoint}}
\put(1030.72,196.14){\usebox{\plotpoint}}
\put(1051.32,198.59){\usebox{\plotpoint}}
\multiput(1054,199)(20.694,1.592){0}{\usebox{\plotpoint}}
\put(1071.94,200.76){\usebox{\plotpoint}}
\multiput(1080,202)(20.684,1.724){0}{\usebox{\plotpoint}}
\put(1092.56,203.09){\usebox{\plotpoint}}
\put(1113.07,206.24){\usebox{\plotpoint}}
\multiput(1118,207)(20.514,3.156){0}{\usebox{\plotpoint}}
\put(1133.58,209.43){\usebox{\plotpoint}}
\put(1153.92,213.52){\usebox{\plotpoint}}
\multiput(1156,214)(20.224,4.667){0}{\usebox{\plotpoint}}
\put(1174.12,218.28){\usebox{\plotpoint}}
\multiput(1181,220)(20.224,4.667){0}{\usebox{\plotpoint}}
\put(1194.31,223.09){\usebox{\plotpoint}}
\put(1214.15,229.20){\usebox{\plotpoint}}
\multiput(1220,231)(19.159,7.983){0}{\usebox{\plotpoint}}
\put(1233.48,236.68){\usebox{\plotpoint}}
\put(1252.53,244.90){\usebox{\plotpoint}}
\multiput(1258,247)(18.275,9.840){0}{\usebox{\plotpoint}}
\put(1271.11,254.07){\usebox{\plotpoint}}
\put(1288.31,265.68){\usebox{\plotpoint}}
\put(1305.04,277.96){\usebox{\plotpoint}}
\put(1320.15,292.15){\usebox{\plotpoint}}
\put(1333.80,307.77){\usebox{\plotpoint}}
\put(1345.97,324.58){\usebox{\plotpoint}}
\put(1357.00,342.16){\usebox{\plotpoint}}
\put(1365.93,360.84){\usebox{\plotpoint}}
\multiput(1372,375)(7.049,19.522){2}{\usebox{\plotpoint}}
\multiput(1385,411)(5.426,20.034){2}{\usebox{\plotpoint}}
\multiput(1398,459)(3.790,20.407){4}{\usebox{\plotpoint}}
\multiput(1411,529)(2.292,20.629){5}{\usebox{\plotpoint}}
\multiput(1423,637)(1.409,20.708){9}{\usebox{\plotpoint}}
\put(1436,828){\usebox{\plotpoint}}
\sbox{\plotpoint}{\rule[-0.400pt]{0.800pt}{0.800pt}}%
\special{em:linewidth 0.8pt}%
\put(1306,722){\makebox(0,0)[r]{$f_0 < f_s$}}
\put(1328,722){\special{em:moveto}}
\put(1394,722){\special{em:lineto}}
\put(176,274){\special{em:moveto}}
\put(189,265){\special{em:lineto}}
\put(201,257){\special{em:lineto}}
\put(214,250){\special{em:lineto}}
\put(227,243){\special{em:lineto}}
\put(240,237){\special{em:lineto}}
\put(252,231){\special{em:lineto}}
\put(265,226){\special{em:lineto}}
\put(278,221){\special{em:lineto}}
\put(291,217){\special{em:lineto}}
\put(303,212){\special{em:lineto}}
\put(316,209){\special{em:lineto}}
\put(329,205){\special{em:lineto}}
\put(341,201){\special{em:lineto}}
\put(354,198){\special{em:lineto}}
\put(367,195){\special{em:lineto}}
\put(380,192){\special{em:lineto}}
\put(392,190){\special{em:lineto}}
\put(405,187){\special{em:lineto}}
\put(418,185){\special{em:lineto}}
\put(431,182){\special{em:lineto}}
\put(443,180){\special{em:lineto}}
\put(456,178){\special{em:lineto}}
\put(469,176){\special{em:lineto}}
\put(481,174){\special{em:lineto}}
\put(494,172){\special{em:lineto}}
\put(507,171){\special{em:lineto}}
\put(520,169){\special{em:lineto}}
\put(532,167){\special{em:lineto}}
\put(545,166){\special{em:lineto}}
\put(558,164){\special{em:lineto}}
\put(571,163){\special{em:lineto}}
\put(583,161){\special{em:lineto}}
\put(596,160){\special{em:lineto}}
\put(609,159){\special{em:lineto}}
\put(621,158){\special{em:lineto}}
\put(634,156){\special{em:lineto}}
\put(647,155){\special{em:lineto}}
\put(660,154){\special{em:lineto}}
\put(672,153){\special{em:lineto}}
\put(685,152){\special{em:lineto}}
\put(698,151){\special{em:lineto}}
\put(711,150){\special{em:lineto}}
\put(723,149){\special{em:lineto}}
\put(736,148){\special{em:lineto}}
\put(749,148){\special{em:lineto}}
\put(761,147){\special{em:lineto}}
\put(774,146){\special{em:lineto}}
\put(787,145){\special{em:lineto}}
\put(800,144){\special{em:lineto}}
\put(812,144){\special{em:lineto}}
\put(825,143){\special{em:lineto}}
\put(838,142){\special{em:lineto}}
\put(851,142){\special{em:lineto}}
\put(863,141){\special{em:lineto}}
\put(876,140){\special{em:lineto}}
\put(889,140){\special{em:lineto}}
\put(901,139){\special{em:lineto}}
\put(914,139){\special{em:lineto}}
\put(927,138){\special{em:lineto}}
\put(940,137){\special{em:lineto}}
\put(952,137){\special{em:lineto}}
\put(965,136){\special{em:lineto}}
\put(978,136){\special{em:lineto}}
\put(991,135){\special{em:lineto}}
\put(1003,135){\special{em:lineto}}
\put(1016,134){\special{em:lineto}}
\put(1029,134){\special{em:lineto}}
\put(1041,133){\special{em:lineto}}
\put(1054,133){\special{em:lineto}}
\put(1067,133){\special{em:lineto}}
\put(1080,132){\special{em:lineto}}
\put(1092,132){\special{em:lineto}}
\put(1105,131){\special{em:lineto}}
\put(1118,131){\special{em:lineto}}
\put(1131,131){\special{em:lineto}}
\put(1143,130){\special{em:lineto}}
\put(1156,130){\special{em:lineto}}
\put(1169,130){\special{em:lineto}}
\put(1181,129){\special{em:lineto}}
\put(1194,129){\special{em:lineto}}
\put(1207,129){\special{em:lineto}}
\put(1220,128){\special{em:lineto}}
\put(1232,128){\special{em:lineto}}
\put(1245,128){\special{em:lineto}}
\put(1258,127){\special{em:lineto}}
\put(1271,127){\special{em:lineto}}
\put(1283,127){\special{em:lineto}}
\put(1296,127){\special{em:lineto}}
\put(1309,126){\special{em:lineto}}
\put(1321,126){\special{em:lineto}}
\put(1334,126){\special{em:lineto}}
\put(1347,126){\special{em:lineto}}
\put(1360,125){\special{em:lineto}}
\put(1372,125){\special{em:lineto}}
\put(1385,125){\special{em:lineto}}
\put(1398,125){\special{em:lineto}}
\put(1411,124){\special{em:lineto}}
\put(1423,124){\special{em:lineto}}
\put(1436,124){\special{em:lineto}}
\end{picture} \\ ~\\
Therefore $h^2 (t)$ decreases much faster than $g^2 (t)$ for $m=1,2$.
Then we can ignore $h$ terms at large $t$. Thus, Eq.(\ref{eq:7}) becomes
\be
(4 \pi )^2 \frac{df}{dt} = \frac{11}{3} f^2 -24g^2 f + 72 g^4 .
\ee
The solution is
\be
f(t) = \frac{24 \pi^2}{11} \beta g^2 (t) \frac{\beta \tan[\frac{\beta}{2c}
\ln(1 + c g_0^2 t )] + \frac{11}{24 \pi^2} \frac{f_0}{g_0^2} - a }
{\beta - \tan[\frac{\beta}{2c}
\ln(1 + c g_0^2 t )] ( \frac{11}{24 \pi^2} \frac{f_0}{g_0^2} - a )}
+ \frac{24 \pi^2}{11} a g^2 (t),
\ee
where
$~~~~~\beta \equiv \sqrt{\frac{1}{\pi^4}
(- \frac{1}{9} m^2 - \frac{10}{24} m + \frac{239}{64})}.$

For $ f_0 > f_s $ and $ f_0 < f_s $ at $ h_0^2 < h_s^2 $,
it is difficult to manipulate the equations analytically. 
Thus, numerical solution is investigated for this case.
The initial $g_0$ is given by $g_0 = 0.650$ \cite{kn:7}.
In the case of $ h_0^2 < h_s^2 $, the $ f(t) $ are similar to a parabola
independent of choice of $f_0$ [Fig.3]. \\ ~\\

\setlength{\unitlength}{0.240900pt}
\ifx\plotpoint\undefined\newsavebox{\plotpoint}\fi
\sbox{\plotpoint}{\rule[-0.200pt]{0.400pt}{0.400pt}}%
\special{em:linewidth 0.4pt}%
\\ ~\\
The solution of $\xi$ in the case of $h_0^2 = h_s^2 $, $ f_0 \neq f_s$
at large $t$ is
\be
\xi = ( \xi_0 - \frac{1}{6}) ( 1 + c g_0^2 t )^\frac{D}{c} + \frac{1}{6}  ,
\ee
where
\be
D \equiv \frac{1}{16\pi^2} [ - \frac{5}{3}
\sqrt{\frac{24}{11} (\frac{69}{16} + 5 m + \frac{4}{3} m^2 ) }
+ 8 (\frac{5}{8} + \frac{1}{3} m ) -12 ].
\ee
For $m=1$ and $2$, 
$\xi \rightarrow \frac{1}{6}$ as $t \rightarrow \infty$,
because $\frac{D}{c}$ is negative [Fig.4].
But the potential gets unstable because $f(t)$ becomes negative at large $t$.
For $h_0^2 < h_s^2$, we can ignore $h^2 (t)$ for large $t$ in Eq.(\ref{eq:8}).
Then, Eq.(\ref{eq:8}) becomes

\be
(4 \pi)^2 \frac{d\xi}{dt} =
(\xi - \frac{1}{6})( \frac{5}{3} f - 12g^2 ) .
\ee
Then, the solution of $\xi$ for large t is
\be
\xi = (\xi_0 - \frac{1}{6}) \exp [ { \ln [ \cos{(\theta + \kappa)}^{\frac{20}{11}}]
+ \ln [(1+ cg_0^2 t )^{- \frac{1}{c} (\frac{6}{11}a + \frac{3}{\pi^2})}]
+ const }] + \frac{1}{6} ,
\ee
where

$~~~~~~~ \theta \equiv \frac{\beta}{2c} \ln(1 + c g_0^2 t ),~~
\kappa \equiv \arctan[{\frac{11}{24\pi^2} \frac{f_0}{g_0^2} + a}].$\\
Therefore, $\xi \rightarrow \frac{1}{6}$, when $t\rightarrow \infty$
for $m=1,2$.

\noindent
\section{Standard Model in Curved Space-Time}

Nowadays, the Standard Model(SM) is considered as the most believable theory.
We may ask whether the SM has the asymptotic conformal invariance in curved
space-time!
We have investigated the question through the SM one-loop RG-equations.

We choose a gauge the 'tHooft-Landau gauge. In this gauge the, $W^{\pm}$,
$Z$ and photon are transverse, and the associate ghosts are massless and
couple only to the gauge fields;
the would be goldstone bosons $G^{\pm},G$ have a common
mass driving from the scalar potential only. Moreover, the gauge parameter is
not renormalized in this gauge, so it does not enter into the RG equations
\cite{kn:7,kn:8}.

The effective potential $V(\phi)$ through one-loop is \cite{kn:7,kn:8}
\ba
V(\phi) &=& \Omega' (\mu, m^2, h, g, g' ) + \frac{1}{2} m^2 \phi^2
+\frac{1}{24} \lambda \phi^4 \nonumber \\
&+& \kappa [ \frac{1}{4} H^2 ( \ln{\frac{H}{\mu^2}} - \frac{3}{2} )
            +\frac{3}{4} G^2 ( \ln{\frac{G}{\mu^2}} - \frac{3}{2} )
            +\frac{3}{2} W^2 ( \ln{\frac{W}{\mu^2}} - \frac{5}{6} )
			 \nonumber \\
           &+&\frac{3}{4} Z^2 ( \ln{\frac{Z}{\mu^2}} - \frac{5}{6} )
            - 3 T^2 ( \ln{\frac{T}{\mu^2}} - \frac{3}{2} )] + ..., .
\ea\\
where
\ba
&\kappa = 16\pi^2,~~~ &H = m^2 + \frac{1}{2} \lambda \phi^2,\\ \nonumber
&T = \frac{1}{2} h^2 \phi^2, &G = m^2 + \frac{1}{6} \lambda \phi^2,\\ \nonumber
&W = \frac{1}{4} g^2 \phi^2, &Z = \frac{1}{4} (g^2 + g'^2 ) \phi^2.
\ea\\
The SM one-loop RG-equations are \cite{kn:7,kn:8}
\ba
16 \pi^2 \frac{dg^2}{dt} &=& - \frac{19}{3}g^4 , \\
16 \pi^2 \frac{dg'^2}{dt} &=& \frac{41}{3} g'^4 , \\
16 \pi^2 \frac{dg_3^2}{dt} &=& - 14 g_3^4  , \\
16 \pi^2 \frac{dh^2}{dt} &=& 9 h^4 - 16 g_3^2 h^2 -\frac{9}{2} g^2 h^2
-\frac{17}{6} g'^2 h^2  , \\
16 \pi^2 \frac{d\lambda}{dt} &=& 4 \lambda^2 + 12 \lambda h^2 -36 h^4
-9 \lambda g^2 \nonumber \\
&-&3 \lambda g'^2 +\frac{9}{4} g'^4 + \frac{9}{2} g^2 g'^2
+ \frac{27}{4} g^4  , \\
16 \pi^2 \frac{d\xi}{dt} &=& (\xi - \frac{1}{6})
(2 \lambda + 6 h^2 - \frac{9}{2} g^2 - \frac{3}{2} g'^2 ),
\ea \\
where $g',~g,~g_3,~h,~\lambda$, and $\xi$ are U(1), SU(2), SU(3), top quark
Yukawa (other Yukawa couplings are ignored), quartic scalar, and
non-minimal couplings respectively.
Especially, $\beta$-functions for gauge couplings are \cite{kn:8}
\ba
\beta_{g'} = \frac{5}{3} g'^3  ( \frac{4}{3} n + \frac{1}{10} ),~~~
\beta_{g} = g^3  ( \frac{4}{3} n - \frac{43}{6} ),
\ea\\
where $n$ is the number of generation ($n=3$ yields the above results). 
The equation for $\xi$ is obtained using the technique of Ref. \cite{kn:1}.

At $\mu=M_z$, the initial values are \cite{kn:7}\\
$g_0 = 0.650$,\\
$g'_0 = 0.358$,\\
$\alpha_3 = 0.10, 0.11, 0.12, 0.13$,\\
$h_0 = 1.17$,\\
$\lambda_0 = \lambda_0$,\\
$\xi_0 = \xi_0$,\\
where we have used the top quark mass $188GeV$ in deciding the initial
value of $h(t)$\cite{kn:9}.

The analytical solution for U(1), SU(2), and SU(3) couplings can be easily
found as follow;
\ba
g'^2 (t) &=& \frac{g_0'^2}{1- \frac{41}{48 \pi^2} g_0'^2 t } , \\
g^2 (t) &=& \frac{g_0^2}{1+ \frac{19}{48 \pi^2} g_0^2 t } , \\
g_3^2 (t) &=& \frac{g_{30}^2}{1+ \frac{7}{8 \pi^2} g_{30}^2 t }.
\ea
Because the other equations seem to be impossible to have analytical solutions,
numerical solutions are investigated. \\ ~\\

\setlength{\unitlength}{0.240900pt}
\ifx\plotpoint\undefined\newsavebox{\plotpoint}\fi
\sbox{\plotpoint}{\rule[-0.200pt]{0.400pt}{0.400pt}}%
\special{em:linewidth 0.4pt}%
 \\ ~\\

Fig.5 is the graph of gauge couplings for $\alpha_3 = 0.13$.
Increasing the input value of $\alpha_3$
causes $h$ to decrease faster as $t$ increases [Fig.6]. \\ ~\\

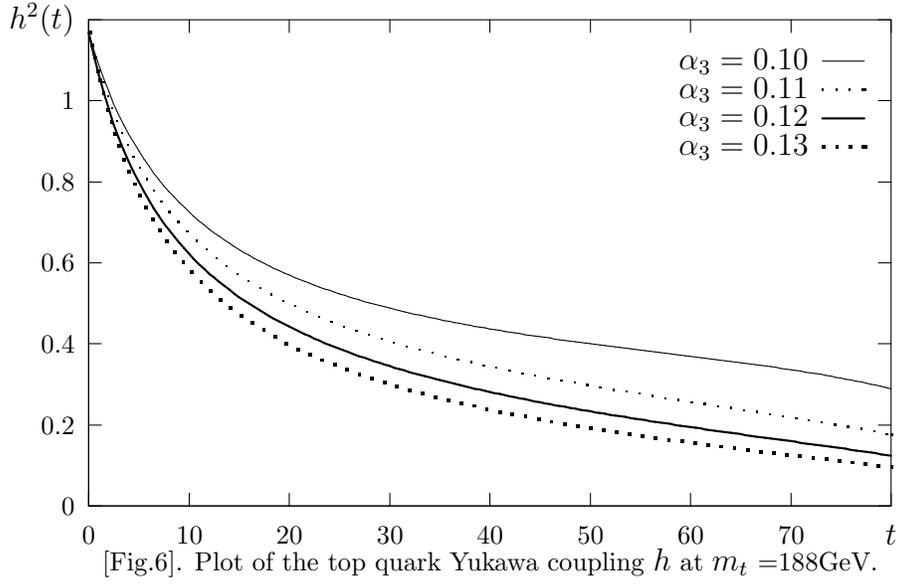
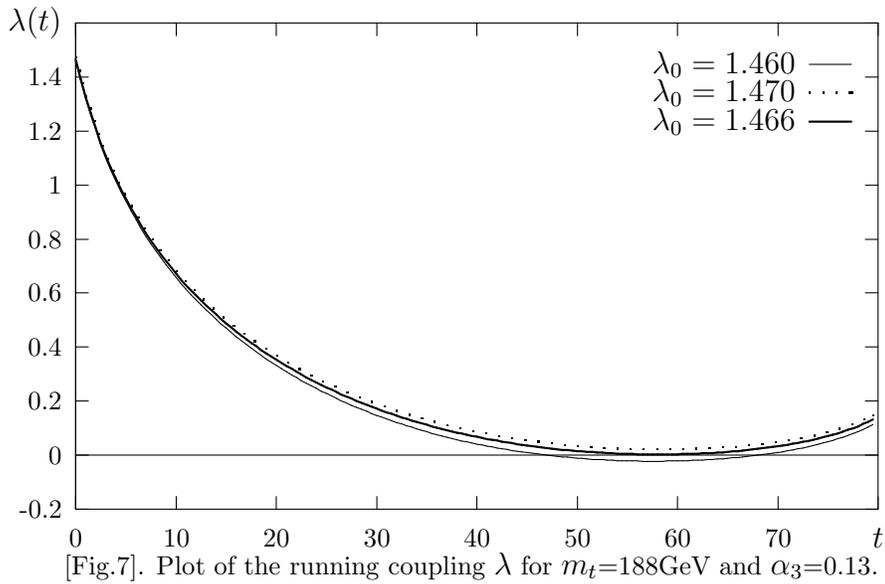
\setlength{\unitlength}{0.240900pt}
\ifx\plotpoint\undefined\newsavebox{\plotpoint}\fi
\sbox{\plotpoint}{\rule[-0.200pt]{0.400pt}{0.400pt}}%
\special{em:linewidth 0.4pt}%
%
 \\ ~\\

The quartic potential coupling $\lambda$ is very sensitive to it's
initial value [Fig.7]. There is a critical $\lambda_0$
such that $\lambda(t)$ is positive at all scale for $\lambda_0 \geq
\lambda_{crit} $, which means the stability of electroweak vacuum.
Requiring the stability of the electroweak vacuum results in the lower limit
on $M_H$ \cite{kn:7,kn:10,kn:11}. In this analysis, we found the
mass of the Higgs as $M_H \geq$178$GeV$ at $\alpha_3 = 0.13$.

The behavior of $\xi(t)$ for SM is similar to a shooting motion [Fig.8].
Standard Model does not have asymptotic conformal invariance at high energy
except the case $\xi_{0}=1/6$. However, some other extensions of SM may have
asymptotic conformal invariance. Moreover, if we consider a grand unified model
at high energy, some GUT models might have the asymptotic conformal invariance 
compared with the results of Ref. \cite{kn:1} where SU(N) asymptotically 
conformal invariant GUT models have been found.

\section{Conclusion}

In some model of the SU(2) gauge theory with scalars and spinors, there
exists the special initial value condition of $h$ and $f$, which makes all
couplings asymptotically free.
In the case of the general solution of RG-equations of couplings,
it is found that the theory has still asymptotic conformal invariance.
In Standard Model, it is found that $\xi$ does not approach to $\frac{1}{6}$ 
asymptotically in general. 
However, if Standard Model has asymptotic conformal invariance, 
$\xi$ should be $\frac{1}{6}$ at all scale (it's unlikely).
Even though there is no asymptotic conformal invariance in Standard Model
generally, but it may happen that other extensions of Standard Model or some 
GUT models may have it \cite{kn:12,kn:13}. \\~\\ \noindent
{\it Acknowledgments:}
We are deeply indebted to S.D. Odintsov for valuable discussions and comments.
This work was supported in part by Hanyang University, KOSEF, and KRF 
through BSRI-2441.

\end{document}